\documentclass[12pt]{article}
\usepackage{graphicx}

\newcommand\pubnumber{DPF2015-82}
\newcommand\pubdate{\today}

\def\Texas{Department of Physics and Astronomy\\
Texas A\&M University, College Station, Texas 77843, USA}

\def\Title#1{\begin{center} {\Large #1 } \end{center}}
\def\Author#1{\begin{center}{ \sc #1} \end{center}}
\def\Address#1{\begin{center}{ \it #1} \end{center}}

\newcommand\pubblock{\rightline{\begin{tabular}{l} \pubnumber\\
         \pubdate  \end{tabular}}}
\newenvironment{Abstract}{\begin{quotation}  }{\end{quotation}}
\newenvironment{Presented}{\begin{quotation} \begin{center} 
             PRESENTED AT\end{center}\bigskip 
      \begin{center}\begin{large}}{\end{large}\end{center} \end{quotation}}

\def\beq{\begin{equation}}
\def\eeq#1{\label{#1}\end{equation}}
\def\eeqn{\end{equation}}

\def\beqa{\begin{eqnarray}}
\def\eeqa#1{\label{#1}\end{eqnarray}}
\def\eeqan{\end{eqnarray}}

\let\bar=\overbar

\def\Dslash{\not{\hbox{\kern-4pt $D$}}}
\def\dslash{\not{\hbox{\kern-2pt $\del$}}}

\def\msb{{\bar{\ssstyle M \kern -1pt S}}}

\begin{document}
\begin{titlepage}
\pubblock

\vfill
\Title{Predicted Higgs-related spin 1/2 particles as a new 
\\
dark matter candidate }
\vfill
\Author{Joshua Stenzel, Johannes Kroll, Minjie Lei, and Roland E. Allen}
\Address{\Texas}
\vfill
\begin{Abstract}
The theory at arXiv:1101.0586 [hep-th] predicts new fundamental spin $1/2$ particles which can be produced in pairs through their couplings to vector bosons or fermions. The lowest-energy of these should have a mass $m_{1/2}$ comparable to the mass $m_h$ of the recently discovered Higgs boson, with $m_{1/2} = m_h$ in the simplest model. These particles should therefore be detectable in collider experiments, perhaps in Run 2 or 3 of the LHC. They cannot decay through any obvious mechanisms in standard physics, making them a new dark matter candidate. In the simplest model, annihilations would produce a well-defined signature with photons, positrons, and excess electrons at about 125 GeV, and the mass would also be well-defined for direct dark matter detection.
\end{Abstract}
\vfill
\begin{Presented}
DPF 2015\\
The Meeting of the American Physical Society\\
Division of Particles and Fields\\
Ann Arbor, Michigan, August 4--8, 2015\\
\end{Presented}
\vfill
\end{titlepage}
\def\thefootnote{\fnsymbol{footnote}}
\setcounter{footnote}{0}

The particle discovered by the ATLAS and CMS collaborations at the LHC is almost certainly a Higgs boson~\cite{ATLAS,CMS,pdg}. After the electron was discovered in 1897, and the photon was introduced by Einstein in 1905, the richness of behavior associated with spin 1/2 fermions and spin 1 gauge bosons emerged
slowly during the following decades. More than a century later, the third kind of Standard Model particle, with spin 0, has finally been discovered, and one should not be completely surprised if some of its implications are
yet to be determined. 

In an earlier paper~\cite{allen-2015}, based on a novel fundamental picture, the following
action for spin 1/2 fermions and scalar bosons was obtained as a low energy
approximation: 
\[
\hspace{-0.2cm}S_{f}+S_{sb}=\int d^{4}x\,\bigg[\psi _{f}^{\dagger }\left(
x\right) \,ie_{\alpha }^{\mu }\,\sigma ^{\alpha }{D}_{\mu }\,\psi _{f}\left(
x\right) -g^{\mu \nu }\left( {D}_{\mu }\phi _{b}\left( x\right) \right)
^{\dagger }{D}_{\nu }\phi _{b}\left( x\right) +F_{b}^{\dagger }\left(
x\right) F_{b}\left( x\right) \bigg]
\]
where 
\[
g^{\mu \nu }=\eta ^{\alpha \beta }e_{\alpha }^{\mu }e_{\beta }^{\nu }\;.
\]
Here $e_{\alpha }^{\mu }$ is the vierbein and $\eta ^{\alpha \beta }$ is the
Minkowski metric tensor in the $(-1,1,1,1)$ convention. The familiar form
for a Lorentz-invariant and supersymmetric action was thus found to follow
automatically from a picture that is intially quite unfamilar. The spin 1/2
fermion fields in $\psi _{f}$, the scalar boson fields in $\phi _{b}$, and
the auxiliary fields in $F_{b}$ span the various physical representations of
the fundamental gauge group, which must be $SO(N)$ in the present theory.
(More precisely, the group is $Spin(N)$, but $SO(N)$ is conventional
terminology.) One unfamiliar feature remains: There is no factor of $e=\left(
-\det g_{\mu \nu }\right) ^{1/2}$ in the integral, and this is related to
the more general fact that the usual cosmological constant vanishes in the
theory of Ref.~\cite{allen-2015}. (The fields $\psi_{f}$, $\phi _{b}$, and $F_{b}$ are required to transform under general coordinate transformations as scalars with weight $1/2$ rather than $0$. Under a Lorentz transformation in the tangent space, they transform in the usual way, as ordinary spinors and scalars respectively.)

At higher energies, including those currently achieved at the LHC, the
theory implies that the above form for the action is no longer a valid
approximation, because internal degrees of freedom can be excited in a
4-component field 
\[
\Phi _{b}=\left( 
\begin{array}{c}
\Phi  \\ 
\Phi _{c}^{\dag }
\end{array}
\right) \; .
\]
This can be written as the inner product of two $N_{g}$-component fields 
$\phi _{b}$ and $\chi _{b}$, where each component of $\phi _{b}$ is a complex
scalar and each component of $\chi _{b}$ is a 4-component bispinor (and
where $N_{g}$ is the number of fields spanning all the physical gauge representations): 
\begin{equation}
\Phi _{b}=\phi _{b}\chi _{b}=\phi _{b}^{r}\chi _{b}^{r}
\end{equation}
with the usual summation over the repeated index $r$. The amplitude of each
component $\Phi _{b}^{r}$ is given by $\phi _{b}^{r}$, and the
\textquotedblleft spin configuration\textquotedblright\ by $\chi _{b}^{r}$.
If $\chi _{b}$ is constant, it is convenient to choose the normalization 
\begin{equation}
\chi _{b}^{r\,\dag }\chi _{b}^{r}=1\quad \mathrm{[no\;sum\;on\;}r\mathrm{]\;.
}
\end{equation}

The more general form of the Lagrangian corresponding to scalar bosons includes the internal
degrees of freedom which are ``hidden'' at low energy. In a locally inertial frame of reference it is
\begin{equation}
\mathcal{L}_{\Phi }=\Phi _{b}^{\dag }\left( x\right) D^{\mu }D_{\mu }\Phi
_{b}\left( x\right) -\frac{1}{2}\left[ \Phi _{b}^{\dag }\left( x\right)
\,S^{\mu \nu }F_{\mu \nu }\,\Phi _{b}\left( x\right) +h.c.\right] 
\end{equation}
where $F_{\mu \nu }$ is the field strength tensor and the $S^{\mu \nu
}=\sigma ^{\mu \nu }/2$ are the Lorentz generators which act on Dirac
spinors. When the second term above is written out explicitly, it involves 
$\phi _{b}^{r\,\dag }\phi _{b}^{r^{\prime }}\chi _{b}^{r\,\dag }\sigma
^{k}\chi _{b}^{r^{\prime }}$ interacting with the \textquotedblleft
magnetic\textquotedblright\ field strengths in $F_{\mu \nu }$ (and is thus
analogous to the interaction of an electron spin with a magnetic field).

Some experimental implications are discussed in Appendix E of Ref.~\cite{allen-2015}. In
particular, the theory predicts new fundamental spin 1/2 particles which can
be produced in pairs through their couplings to vector bosons or fermions.
The lowest-energy of these should have a mass $m_{1/2}$ comparable to the
mass $m_{h}$ of the recently discovered Higgs boson, with $m_{1/2}=m_{h}$ in
the simplest model.

There are two unconventional features in the Lagrangian $\mathcal{L}_{\Phi }$: Each field 
$\Phi _{b}^{r}$ has four components rather than one, and there is a second
term involving the gauge field strengths $F_{\mu \nu }$. One can read off
the general Feynman-diagram vertices for virtual and real processes from the
interactions in each term. These are relevant for all the $\Phi _{b}^{r}$
that correspond to scalar boson fields in standard physics, but let us now
focus on the one $\Phi _{h}$ that corresponds to a single neutral Higgs
field.

The vacuum expectation value of $\Phi _{h}$ has the form 
\[
\left\langle \Phi _{h}^{0}\right\rangle =\frac{v}{\sqrt{2}}\left( 
\begin{array}{c}
1 \\ 
0 \\ 
0 \\ 
1
\end{array}
\right) \;.\;
\]
In Ref.~\cite{allen-2015} it is shown that the condensate then has zero angular momentum and
also no coupling to the gauge fields beyond that in the Standard Model
(since the second term in $\mathcal{L}_{\Phi }$ vanishes when the internal
degrees of freedom in $\Phi _{h}$ are not excited).

The simplest model for excitations of $\Phi _{h}$ has a mass term Lagrangian 
\[
\mathcal{L}_{h}^{\mathrm{mass}}=m_{h}^{2}\,\left( \Delta \Phi _{h}\right)
^{\dag }\Delta \Phi _{h}\;.
\]
When the internal degrees of freedom are not excited, so that $\Delta \Phi
_{h}=h\chi _{0}$ with $\chi _{0}^{\dag }\chi _{0}=1$, the mass term is 
$m_{h}^{2}\,h^{2}$ (for $h$ real). I.e., $m_{h}$ is the mass of the scalar Higgs boson.

For a spin 1/2 excitation with the form 
\[
\Delta \Phi _{h}=\left( 
\begin{array}{c}
h_{+} \\ 
0
\end{array}
\right) \quad \mathrm{or}\quad \Delta \Phi _{h}=\left( 
\begin{array}{c}
0 \\ 
h_{-}
\end{array}
\right) 
\]
we obtain 
\[
\mathcal{L}_{+}^{\mathrm{mass}}=m_{h}^{2}\,h_{+}^{\dag }h_{+}\quad 
\mathrm{or}\quad \mathcal{L}_{-}^{\mathrm{mass}}=m_{h}^{2}\,h_{-}^{\dag }h_{-}\;.
\]
In other words, in the simplest model the spin 1/2 particles $h_{+}$ and 
$h_{-}$ have the same mass $m_{h}$ as the scalar Higgs boson $h$. More
generally, the masses $m_{1/2}$ of these particles should be comparable to 
$m_{h}$. A suggestive 
analogy is s-wave superconductivity, where there are single-particle excitations, two-particle excitations, and ``Higgs
mode''  excitations with minimum energies $\Delta $, $2\Delta $, and $2\Delta $ respectively.

According to the spin-statistics theorem, spin $1/2$ bosonic excitations are
impossible, but the requirements of this theorem are not satisfied in this
one specific context, since $\mathcal{L}_{\Phi }$ is not fully Lorentz
invariant: It is invariant under a rotation, but not a Lorentz boost with
respect to the original (cosmological) coordinate system. The present theory
is, however, fully Lorentz invariant (as well as initially supersymmetric)
if the internal degrees of freedom in $\Phi _{b}$ are not excited  -- and
these excitations can be observed only at the high energies that are now
becoming available. Furthermore, the extremely weak virtual effects of these
excitations are irrelevant to the many existing tests of Lorentz invariance,
which probe those phenomena in various areas of physics and astrophysics
where the present theory is fully Lorentz invariant.

The spin $1/2$ excitations of $\Phi _{b}$ can
be produced in pairs through the coupling to gauge boson fields in $\mathcal{L}_{\Phi }$ -- for
example, by the coupling to virtual or real Z and W bosons. In addition, the
Higgs-related spin $1/2$ particles should have the same basic Yukawa
couplings to fermions as a Higgs boson, since $\Phi _{h}=\phi _{h}\chi _{h}$. 

Once a lowest-mass particle of this kind has left the region where it was created, it is unable to decay through Standard-Model mechanisms without
violating lepton number or baryon number conservation, since the net decay 
products must have angular momentum $1/2$. This implies that these (weakly-interacting) 
particles are dark matter candidates. 
In the simplest model described above, annihilations would produce a well-defined signature with photons, positrons, and excess electrons at about 125 GeV, and the mass would also be well-defined for direct dark matter detection.

\end{document}